\documentclass[prl,twocolumn]{revtex4}
\usepackage{graphicx}% Include figure files
\usepackage{dcolumn}% Align table columns on decimal point
\usepackage{amsmath,amsbsy,amssymb,graphics}
\usepackage{bm}% bold math
\usepackage{color}

\begin{document}

\preprint{APS/123-QED}

\title{
Itinerant vs Localized Heavy-Electron Magnetism
}

\author{Shintaro Hoshino and Yoshio Kuramoto}
\affiliation{Department of Physics, Tohoku University, Sendai 980-8578, Japan}

\date{\today}

\begin{abstract}
It is demonstrated that itinerant-localized transition of heavy electrons
occurs inside the magnetically ordered phase of the Kondo-Heisenberg lattice.  
The phase diagram and electronic structure are derived by means of the continuous-time quantum Monte Carlo combined with the dynamical mean-field theory.
Around the itinerant-localized transition, 
nearly flat bands appear on the Fermi surface with almost vanishing quasi-particle renormalization factor.
At the same time,  there emerges a strong local magnetic fluctuation with a minute energy scale.
Considering both antiferromagnetic and ferromagnetic Heisenberg interactions, coherent understanding is achieved on 
rich phase diagrams observed in
CeRh$_{1-x}$Co$_x$In$_5$, CeRu$_2$(Si$_x$Ge$_{1-x}$)$_2$, UGe$_2$ and CeT$_2$Al$_{10}$ (T=Fe,Ru,Os).

\end{abstract}

\pacs{Valid PACS appear here}% PACS, the Physics and Astronomy
                             % Classification Scheme.
\maketitle

\newcommand{\diff}{\mathrm{d}}
\newcommand{\imag}{\mathrm{Im}\,}
\newcommand{\real}{\mathrm{Re}\,}
\newcommand{\trace}{\mathrm{Tr}\,}
\newcommand{\imu}{\mathrm{i}}

%main text
It has been a long-standing problem whether and how nearly localized $f$ electrons change their character depending on the environment. 
The localized $f$ electrons tend to form magnetically ordered states
with intersite exchange interactions,
 while they acquire the itinerancy by the Kondo effect resulting in a heavy Fermi liquid.
Intriguing phenomena such as quantum criticality and unconventional ordered states have been observed in the competing region, and hence understanding the distinction or duality between itinerant and localized characters is a fundamental and important issue in strongly correlated electron systems.

The difference between itinerant and localized characters appears in the Fermi surface;  itinerant $f$ electrons contribute to the Fermi volume, while localized ones do not.
It has been shown for the Kondo lattice in large dimensions\cite{hoshino10} that
$f$ electrons remain itinerant at the magnetic quantum critical point (QCP).
In this paper, therefore, we focus on possible change of $f$-electron characters {\it inside magnetically ordered states.}
Experimentally,  CeRh$_{1-x}$Co$_x$In$_5$ \cite{settai07} shows, for example, 
that the Fermi surface at $x=0$ resembles the case for localized $f$ electrons, 
while at $x\simeq 0.4$ which is deep inside the antiferromagnetic (AFM) phase \cite{ohira-kawamura07, goh08, sutherland09},
the Fermi surface corresponds to the itinerant one. 
The itinerant-localized transition (ILT) in the ordered state has also been found in CeRu$_2$(Si$_x$Ge$_{1-x}$)$_2$ accompanying the magnetic transition between ferromagnetic (FM) and AFM states \cite{matsumoto11}.
Furthermore, drastic Fermi-surface reconstruction inside the FM phase has been observed in a heavy-electron superconductor UGe$_2$ \cite{aoki12, saxena00, settai02}.
Thus 
it is desirable to explain these behaviors in a unified manner, which   
should lead to deeper understanding of the itinerant-localized duality of electrons. 
Obviously it is necessary to go beyond the celebrated 
Doniach-type phase diagram.

Many theoretical attempts have already been made
to explain the ILTs \cite{pepin07, paul07, martin08, otsuki09, si10, hoshino10}.
The presence of ILTs inside the ordered phase has been pointed out within Gutzwiller-type variational
\cite{watanabe07}, and mean-field 
\cite{isaev13} approximations.
These approximations, although simple, are unable to deal with the duality of electrons; 
the itinerant and localized behaviors depend on the energy scale.
On the other hand, numerical approaches must be accurate enough
to deal with a minute energy scale that can be even smaller than the Kondo temperature in the quantum critical region.  
In this paper, we use the continuous-time quantum Monte Carlo \cite{rubtsov05, gull11}
combined with the dynamical mean-field theory (DMFT) \cite{kuramoto85, georges96} which takes full account of local correlations, and becomes exact in the limit of high dimensions.
As we shall show in the following,  the ILT of $f$ electrons 
is controlled by sign and magnitude of the Heisenberg exchange.
The critical temperature can be tuned to minute, which makes the first-order transition to
a quantum critical transition.
There appears considerable asymmetry on both sides of the ILT that can be probed by a 
large effective mass and enhanced local magnetic susceptibility.
The resultant phase diagram leads to a coherent understanding for the experimentally observed ILTs explained above.

In order to discuss the ILT, we consider the following Kondo-Heisenberg lattice:
\begin{align}
{\cal H} = \sum_{\bm{k} \sigma} \xi_{\bm k} c_{\bm k \sigma}^\dagger c_{\bm k \sigma}
\hspace{-0.5mm}+\hspace{-0.5mm}
 J_{\rm K} \sum_i \bm{S}_i \cdot \bm{s}_{{\rm c}i}
\hspace{-0.5mm}+\hspace{-0.5mm}
 \frac{J_{\rm H}}{z} \sum_{(ij)} \bm{S}_i \cdot \bm{S}_j
,
\end{align}
where $\xi_{\bm k} = \varepsilon_{\bm k} - \mu$ with $\mu$ being the chemical potential.
The first and second terms are the kinetic energy of conduction ($c$) electrons and local Kondo interaction, respectively.
The Heisenberg interaction in the third term realizes the simplest localized magnetism, and is incorporated in the DMFT as the mean field.
The summation with respect to $(ij)$ is taken over pairs of the
nearest neighbor sites with the coordination number $z$.
We take a bipartite lattice with the semi-circular density of states
$D(\varepsilon) = (4/\pi W^2) \sqrt{W^2 - 4\varepsilon^2}$.
The band width
$W=1$ 
is the unif of energy.
Even if the inter-site interaction $J_{\rm H}$ is small as compared to $J_{\rm K}$, 
$J_{\rm H}$ can still be important since it competes with the Kondo energy $T_{\rm K} \propto \exp [-1/D (\mu) J_{\rm K} ]$ instead of $J_{\rm K}$, as will be demonstrated later.
Throughout this paper, the number of $c$ electrons is fixed as $n=0.95$ per site.

\begin{figure}[t]
\begin{center}
\includegraphics[width=70mm]{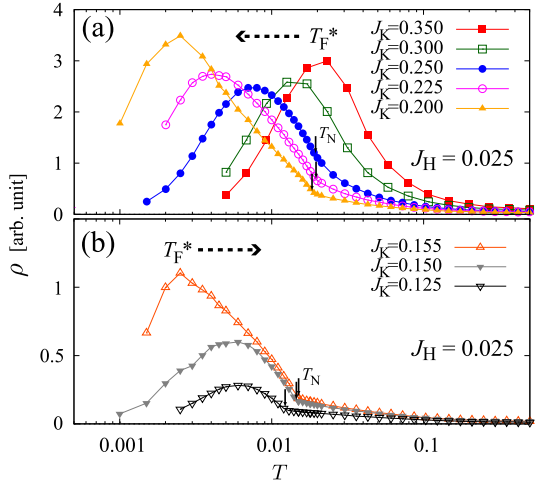}
\caption{
(Color online)
Temperature dependence of the electrical resistivity for (a) $J_{\rm K} > J_{\rm c2} \simeq 0.16$ and (b) $J_{\rm K} < J_{\rm c2}$,
where $J_{\rm c2}$ gives the ILT as discussed in the text.
The dotted arrow shows the change of $T_{\rm F}^*$ with decreasing $J_{\rm K}$.
}
\label{fig_resis}
\end{center}
\end{figure}

Let us first derive the characteristic energy scale 
from temperature dependence of the electrical resistivity $\rho(T)$.
We use the formula
\begin{align}
\rho(T)^{-1} = \alpha \lim_{\omega \rightarrow 0} {\imag \Pi(\bm{q}=\bm{0},\omega)}/{\omega}
. \label{eq_resis}
\end{align}
Here $\Pi(\bm{q},\omega)$ is the current-current correlation function and $\alpha$ is a constant.
Since the local vertex correction in the DMFT does not contribute to the conductivity \cite{georges96}, we evaluate a simple particle-hole bubble.

\begin{figure}[t]
\begin{center}
\includegraphics[width=75mm]{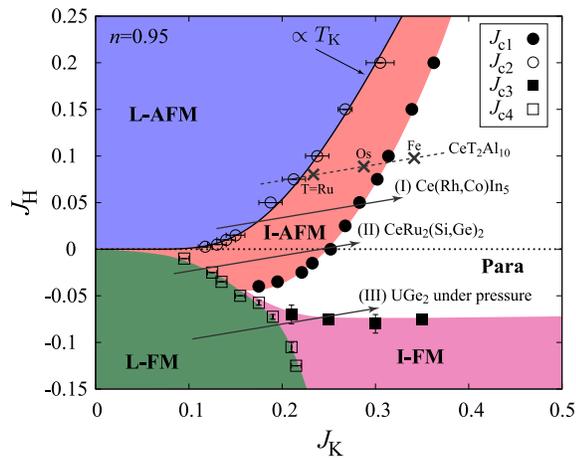}
\caption{
(Color online)
Phase diagram of the Kondo-Heisenberg lattice in the 
$J_{\rm H}$-$J_{\rm K}$ plane at $T=0.001$.
The abbreviations L- and I- mean localized and itinerant, respectively.
At the critical value $J_{\rm K} = J_{\rm c2}$ for the ILT,  the corresponding $J_{\rm H}$
mostly scales with the Kondo temperature $T_{\rm K}$.
}
\label{fig_gs}
\end{center}
\end{figure}

Figure~\ref{fig_resis} shows the temperature dependence of resistivity for the various values of $J_{\rm K}$.
We fix the Heisenberg exchange as $J_{\rm H} = 0.025$
except for Fig.~\ref{fig_gs}.
For $J_{\rm K} = 0.350$ shown in Fig.~\ref{fig_resis}(a), 
the system remains paramagnetic down to zero temperature.  The resistivity at high $T$
increases with decreasing $T$ due to the impurity-like Kondo effect.
On the other hand, the resistivity shows the metallic behavior at low temperatures reflecting the development of coherence.
The characteristic temperature $T_{\rm F}^* \simeq 0.02$, 
which gives the peak of $\rho(T)$, corresponds to 
the effective Fermi temperature
below which the heavy Fermi liquid is formed.
As seen in Fig. \ref{fig_resis}(a), $T_{\rm F}^*$
decreases with decreasing $J_{\rm K}$.

With $J_{\rm K} < J_{\rm c1}\simeq 0.27$, the system undergoes an AFM transition at the N\'{e}el temperature $T_{\rm N}$ as indicated by arrows in Fig.~\ref{fig_resis}.
The resistivity shows almost no anomaly at the transition point with $J_{\rm K} = 0.250$.
The kink at $T_{\rm N}$ becomes clearer for smaller $J_{\rm K}$.
Contrary to the case with larger $J_{\rm K}$ shown in the upper panel (a), the effective Fermi temperature {\it increases} with decreasing $J_{\rm K}$ for 
smaller Kondo interactions 
as shown in Fig.~\ref{fig_resis}(b).

\begin{figure}[t]
\begin{center}
\includegraphics[width=75mm]{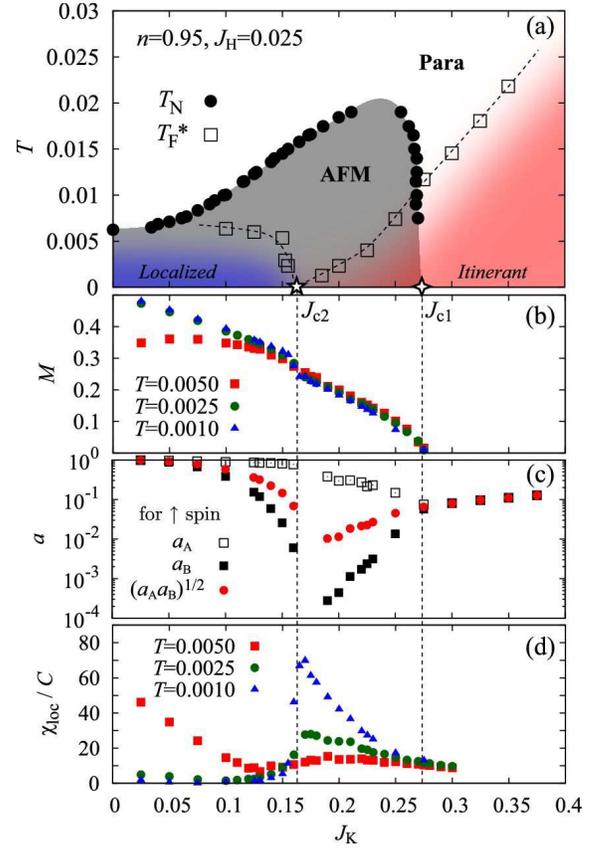}
\caption{
(Color online)
(a) Temperature $(T)$ vs Kondo interaction $(J_{\rm K})$ phase diagram for $J_{\rm H}=0.025$.
The $J_{\rm K}$ dependences of (b) magnetic moment, (c) renormalization factors and (d) local magnetic susceptibility are also shown.
}
\label{fig_phase}
\end{center}
\end{figure}

\begin{figure}[t]
\begin{center}
\includegraphics[width=60mm]{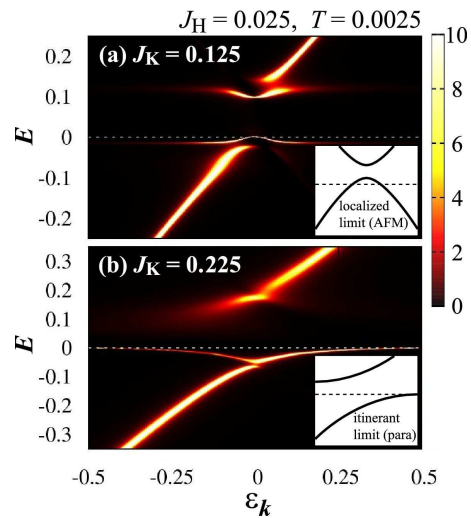}
\caption{
(Color online)
Single-particle spectrum $A_{\sigma}(\varepsilon_{\bm k}, E)$ at $T=0.0025$ for (a) $J_{\rm K} = 0.125$ and (b) $J_{\rm K} = 0.225$.
The insets in (a) and (b) show the schematic illustrations for spectra in the localized (AFM) and itinerant (Para) limits, respectively.
}
\label{fig_spect}
\end{center}
\end{figure}

We plot in Fig.~\ref{fig_phase}(a) the N\'{e}el temperature $T_{\rm N}$ and Fermi temperature $T_{\rm F}^*$ as functions of $J_{\rm K}$.
Note that $T_{\rm F}^*$ is finite 
at the magnetic QCP with $J_{\rm K} = J_{\rm c1} \simeq 0.27$, 
but it becomes almost zero at $J_{\rm K} = J_{\rm c2} \simeq 0.16$ inside the AFM phase.
This point actually separates the itinerant and localized behaviors of $f$ electrons, and is called the ILT point.
More details will be discussed later around Fig.~\ref{fig_spect}.
From the localized side,
$T_{\rm F}^*$ 
tends to zero toward $J_{\rm c2}$
much more rapidly as compared to the itinerant side.  
% The asymmetry is discussed in more detail later.

Figures \ref{fig_phase}(b), (c), (d) show relevant
physical quantities around the ILT.
The spontaneous magnetic moment $M=|\langle S_i^z + s_{{\rm c}i}^z \rangle|$ in (b)
shows a kink at the ILT point at $T=0.001$, while it becomes smooth at $T=0.005$.
This 
behavior reflects the first order transition at 
temperatures lower than $T=0.001$.
Namely the critical end point where the first-order transition line terminates should be located 
at certain $T < 0.001$.
We have not observed the ILT at half filling with $n = 1$ where the system is insulating.
Hence, the temperature at the critical end point should tend to zero as $n \rightarrow 1$.

Figure \ref{fig_phase}(c) shows the renormalization factors 
$a_{\lambda\uparrow}$, which is defined by
\begin{align}
a_{\lambda \sigma} =  \lim_{\varepsilon \rightarrow 0}  \left( 1 - \frac{\partial \Sigma_{\lambda \sigma}}{\partial \varepsilon} \right) ^{-1}
,
\end{align}
for 
sublattice components $\lambda = {\rm A}, {\rm B}$ and 
spin $\sigma = \uparrow, \downarrow$.
Here $\Sigma_{\lambda \sigma} (\varepsilon)$ is the local self energy of $c$ electrons.
Note that the relations $\Sigma_{{\rm A}\uparrow} = \Sigma_{{\rm B}\downarrow}$ and $\Sigma_{{\rm A}\downarrow} = \Sigma_{{\rm B}\uparrow}$
hold in the AFM phase.
We evaluate $a_{\lambda \sigma}$
numerically from the low-energy behavior of the self energy at sufficiently low temperatures.
For example, we take $T=0.001$ for $J_{\rm K} = 0.225$ and $T=0.0025$ for $J_{\rm K}=0.35$.
In the paramagnetic state, $a_{\rm A\uparrow}$ ($= a_{\rm B\uparrow}$) decreases with decreasing $J_{\rm K}$ reflecting the reduction of $T_{\rm K}$.

In the AFM phase, 
$a_{\rm B\uparrow}$ 
is strongly reduced near the ILT point, while $a_{\rm A\uparrow}$ is enhanced toward the ILT from the itinerant side.  
Note that we have the relation $n_{\rm A\uparrow} > n_{\rm B\uparrow}$ with $n_{\lambda\uparrow}$ being the number of $c$ electrons with $\uparrow$ spin at sublattice $\lambda$.
In the localized side, $a_{\rm A\uparrow}$ is almost unity.
Namely, the large renormalization 
takes place only for $c$ electrons with the minority spin at each site.
As shown elsewhere \cite{hoshino2013}, 
the geometric mean $\sqrt{a_{\rm A\uparrow}a_{\rm B\uparrow}}$ contributes to the specific heat coefficient for simple bipartite lattices.
This quantity is also plotted in Fig.~\ref{fig_phase}(c), and behaves in a manner similar to $a_{\rm B\uparrow}$.
The behavior of $\sqrt{a_{\rm A\uparrow}a_{\rm B\uparrow}}$ is consistent with $T_{\rm F}^*$ in Fig.~\ref{fig_phase}(a) in that the energy scale becomes smaller toward $J_{\rm c2}$, and has the 
asymmetry between itinerant and localized sides.

The magnetic response 
near the ILT shows a peculiar enhancement 
as shown in Fig.~\ref{fig_phase}(d).
We 
derive the local magnetic susceptibility defined by
\begin{align}
\chi_{\rm loc} = \lim_{H_i \rightarrow 0}  \frac{\langle S_i^z \rangle}{H_i}
,
\end{align}
where $H_i$ is the local magnetic field applied only at site $i$.
Note that $\chi_{\rm loc}$ is 
independent of the site index even in the AFM phase.
The results are shown in Fig.~\ref{fig_phase}(d).
For $J_{\rm K} \lesssim 0.15$ and $
J_{\rm K}\gtrsim 0.25 $, 
$\chi_{\rm loc}$ is almost the same for 
$T=0.0025$ and $T=0.0010$
shown in the figure.
The smallness of $\chi_{\rm loc}$ in the small $J_{\rm K}\ (\lesssim 0.15)$ region is due to the large magnetic moment, while 
in the large $J_{\rm K}\ (\gtrsim 0.25)$ region, the smallness is due
to the formation of Kondo singlets.

Near the ILT point,
the local magnetic susceptibility $\chi_{\rm loc}$ continuously increases with decreasing temperature, even though the magnetic moment is already large
at $J_{\rm c2}$ as shown in Fig.~\ref{fig_phase}(b).
The large fluctuation in the itinerant regime 
rapidly decreases 
on entering the localized side.
We have checked (not shown) that
uniform and staggered susceptibilities 
also show the rapid change at $J_{\rm c2}$.
The asymmetry between itinerant and localized sides is similar to
$T_{\rm F}^*$ as shown in Fig.~\ref{fig_phase}(a),

This anomalous magnetic fluctuation {\it deep inside} the magnetic phase originates from the formation of extremely narrow electronic bands caused by the Kondo effect.
The distinction between the AFM phases above and below $J_{\rm c2}$ is most clearly
visible in the single-particle spectral function defined by
\begin{align}
A_{\sigma}(\varepsilon_{\bm k}, E) = \frac{1}{f(E)} \int_{-\infty}^{\infty} \frac{\diff t}{2\pi}
\langle c_{\bm{k}\sigma}^\dagger (t) c_{\bm{k}\sigma} \rangle e^{-\imu E t}
, \label{eq_def_spect}
\end{align}
where ${\cal O}(t) = e^{\imu {\cal H} t} {\cal O} e^{-\imu {\cal H} t}$ and $f(E) = 1/ (e^{E/T} + 1)$.
The $\bm k$ dependence of $A_{\sigma}(\varepsilon_{\bm k}, E)$ in the DMFT enters only through $\varepsilon_{\bm k}$ since the self energy has no wave-vector dependence.
In the present notation, the Brillouin zone center and corner are located at $\varepsilon_{\bm k} = -1/2$ and $\varepsilon_{\bm k} = 1/2$, respectively.
The energy $\varepsilon_{\bm k} = 0$
gives the center of band folding by the cell doubling.
Namely, $\varepsilon_{\bm k} = 0$ gives the boundary of the new Brillouin zone (BZ).
Note that the spectrum (\ref{eq_def_spect}) is 
independent of spin in the AFM phase.

Figure \ref{fig_spect} shows the single-particle spectrum at $T=0.0025$.
In the localized regime shown in Fig.~\ref{fig_spect}(a), the Fermi surface is located near the new BZ boundary.
The location of the minimum energy gap is
the same as that in the localized limit $J_{\rm K}\rightarrow 0$,  as illustrated in the inset.
In the itinerant regime,
 on the other hand, 
the Fermi surface is almost the same as in the paramagnetic heavy Fermi liquid as shown in Fig.~\ref{fig_spect}(b) and its inset.
Hence this state is classified as itinerant.
In the itinerant regime, a very flat band appears at the Fermi level, while in the localized side the band is more dispersive at $E=0$
as shown in Fig.~\ref{fig_spect}(a).
This is related to the 
difference between the main components of 
the Fermi surface: mainly $c$ electrons in the localized regime, but mainly $f$ electrons in the itinerant side.
It is this nearly flat band that causes the large magnetic fluctuation as seen in Fig.~\ref{fig_phase}(d). 
We note that the present model has no charge degrees of freedom for $f$ electrons.
Hence it is appropriate to interpret 
the emergent flat band 
in terms of coherent interaction of Kondo peaks, 
rather than direct $cf$ hybridization.

As shown in Fig.~\ref{fig_spect}(a), 
a partly hybridized band structure is observed even for $J_{\rm K} < J_{\rm c2}$, which belongs to our localized regime.
This situation makes it necessary to define precisely the meaning of localized electrons.  
We rely on the character of the Fermi surface, since this distinction is most practical and accessible by experiment.
Indeed, the ILT separates the two regimes with rather different physical properties as seen in Figs.~\ref{fig_resis} and \ref{fig_phase}.

We have performed similar calculations for various values of $J_{\rm K}$ and $J_{\rm H}$, and obtained 
the phase diagram of the Kondo-Heisenberg lattice.
Figure \ref{fig_gs} shows the result
 at $T=0.001$, which we regard as close
to the ground state.
Note, however, that we have not considered superconductivity since the DMFT cannot deal with pairings other than the $s$-wave.
The label L or I in the figure means the localized or itinerant magnetism, respectively.
The blank circles and squares
show the points where a discontinuous change takes place.
At $J_{\rm H}=0$, 
which corresponds to
the ordinary Kondo lattice, the AFM phase has an itinerant ground state
regardless of the value of $J_{\rm K}$ ($< J_{\rm c1}$).
For small Kondo interaction, however, the itinerant AFM is fragile and easily turns into the localized one beyond
a critical $J_{\rm H}\ (>0)$.
The critical value of $J_{\rm H}$ for the ILT with given $J_{\rm K}$
scales well with $T_{\rm K}$ as shown in Fig.~\ref{fig_gs}.
This clearly shows that the ILT in the AFM phase occurs if the Kondo temperature exceeds the Heisenberg interaction.

Next let us discuss the FM state for $J_{\rm H} < 0$.
For small enough $|J_{\rm H}|$, the 
localized ferromagnetism (L-FM) changes into itinerant AFM with increasing $J_{\rm K}$, which is indicated by $J_{\rm c4}$ in Fig.~\ref{fig_gs}.
For larger Heisenberg interaction with $|J_{\rm H}| \gtrsim 0.07$,
the transition occurs between I- and L-FMs. 
There is another phase transition between I-FM and paramagnetic phases, and the corresponding value
 is indicated by $J_{\rm c3}$ in Fig.~\ref{fig_gs}.
The I-FM phase seems to persist up to large Kondo couplings,
and can be understood 
as the spin selective Kondo insulator \cite{peters12}.
This state is interpreted as a mixture of fully polarized localized spins and local Kondo singlets.
Here $c$ electrons with the minority spin are insulating, while those with the majority spin are metallic.
We note that the I-AFM cannot be understood in a manner similar to the I-FM, since the I-AFM phase is not connected to $J_{\rm K}\rightarrow \infty$.

Unlike the AFM, the itinerant and localized FMs {\it can} be connected  either discontinuously or continuously.
In the present case with $n=0.95$, 
the L-FM changes to the I-FM by a first-order transition.
As $n$ decreases, however, the first-order transition changes into a continuous crossover. 
We have actually observed a
continuous change for $n \lesssim 0.5$.
In Fig.~\ref{fig_gs}, we have defined the L-FM state as the phase which is continuously connected to the localized limit with $J_{\rm K} = 0$.

Let us discuss relevance of our results to 
magnetic properties in real systems at low temperatures.
With increasing pressure, the Kondo interaction will become larger in Ce and U systems.
The arrows in Fig.~\ref{fig_gs} show this trend for $J_{\rm K}$.  On the other hand, it is not so simple to simulate the pressure dependence of $J_{\rm H}$.  Hence angle of the arrows do not have a microscopic basis.
Let us first begin with the arrow (I) that 
simulates the behavior in CeRh$_x$Co$_{1-x}$In$_5$ \cite{goh08}. Namely, decrease of $x$ corresponds to the chemical pressure, and 
we expect the L-AFM to change into I-AFM and then to become paramagnetic.
The magnetic structure in the I-AFM 
seems different from that in the L-AFM as observed in this compound \cite{yokoyama08}.
With fine tuning of parameters, $J_{\rm c1}$ can be made very close to $J_{\rm c2}$. Then the ILT occurs very close to the QCP, which has been seen in CeRhIn$_5$ under pressure \cite{shishido05, settai07}.

Next we turn to the arrow (II) in Fig.~\ref{fig_gs} that simulates  CeRu$_2$(Si$_x$Ge$_{1-x}$)$_2$ \cite{matsumoto11}.  
CeRu$_2$Ge$_2$ with $x=0$ is a localized ferromagnet, which changes into itinerant antiferromagnet with increasing Si content, and finally becomes paramagnetic at $x=1$.
If $J_{\rm H}$ is more ferromagnetic, we obtain the arrow
(III) in Fig.~\ref{fig_gs} that simulates
UGe$_2$ under pressure, where the Fermi surface changes as the localized ferromagnetism changes into the itinerant one
inside the magnetic phase \cite{aoki12}.

Another interesting example is found in Kondo insulators with magnetic order.  Namely, 
CeRu$_2$Al$_{10}$ is a heavy-electron antiferromagnet \cite{nishioka09, robert10}, 
which is regarded as located close to the I-AFM and L-AFM boundary as illustrated
in Fig.~\ref{fig_gs}.
Slight substitution of Ru by Rh
changes its itinerant character into a localized one \cite{kondo13}.
Instead of Ru, we obtain 
CeOs$_2$Al$_{10}$ \cite{nishioka09}, which has a larger Kondo interaction, and should be  
located near $J_{\rm c1}$.
On the other hand, CeFe$_2$Al$_{10}$ \cite{nishioka09} has the strongest $cf$ interaction among the CeT$_2$Al$_{10}$ family with T = Ru, Os, Fe and remains paramagnetic down to lowest accessible temperatures.
The crosses in Fig.~\ref{fig_gs} indicate locations of 
these compounds.
Note that the values of $J_{\rm H}$ are uncertain and not important for sufficiently positive $J_{\rm H}$.

Finally, we discuss implication of the strong local magnetic fluctuations  
near the ILT as shown in Fig.~\ref{fig_phase}(d). 
The entropy associated with the fluctuations should be removed at sufficiently low temperatures.
It is tempting to interpret the superconductivity 
in CeRh$_x$Co$_{1-x}$In$_5$ \cite{goh08}
as a result of releasing the entropy.
In the boundary of L-FM and I-FM, the local fluctuation becomes 
also strong especially in the itinerant side.
Hence 
the superconductivity observed in UGe$_2$ \cite{kabeya09} seems 
understandable by the same mechanism.

In conclusion, we have identified the ILT inside ordered states of the Kondo-Heisenberg lattice by means of the DMFT.
An extremely small energy scale appears near the transition point.
It has been proposed that the observed phase diagrams in CeRh$_x$Co$_{1-x}$In$_5$, CeRu$_2$(Si$_x$Ge$_{1-x}$)$_2$,
UGe$_2$ and CeT$_2$Al$_{10}$ can be understood by the present model in a unified way.

%acknowledgement
We are grateful to Noriyuki Kabeya, Yuji Matsumoto and Hiroshi Tanida for fruitful discussions.
SH acknowledges the financial support from JSPS.
This work was partly supported by a Grant-in-Aid for Scientific Research on Innovative Areas ``Heavy Electrons" (No 20102008) of MEXT.
The calculations were partly performed on supercomputer in ISSP, University of Tokyo.

\end{document}